\documentclass[]{spie}  

\usepackage{amsmath,amsfonts,amssymb}
\usepackage{graphicx}
\usepackage[colorlinks=true, allcolors=blue]{hyperref}

\usepackage{amsfonts}
\usepackage{amsmath}
\usepackage{graphicx}      
\usepackage{float} 
\usepackage{psfrag} 
\usepackage{mathtools}
\usepackage{url}
\usepackage{cite}

\title{Data-driven Estimation, Tracking, and System Identification of Deterministic and Stochastic Optical Spot Dynamics}

\author[a]{Aleksandar Haber}
\author[a]{Michael Krainak}

\affil[a]{Relative Dynamics Inc., 14400 Sweitzer Lane, STE. 125,  Laurel, MD 20707, USA }

\authorinfo{Further author information: (Send correspondence to A. H.)\\A. H.: E-mail: aleksandar.haber@gmail.com}

\begin{document} 
\maketitle

\begin{abstract}
Stabilization, disturbance rejection, and control of optical beams and optical spots are ubiquitous problems that are crucial for the development of optical systems for ground and space telescopes, free-space optical communication terminals, precise beam steering systems, and other types of optical systems. High-performance disturbance rejection and control of optical spots require the development of disturbance estimation and data-driven Kalman filter methods. Motivated by this, we propose a unified and experimentally verified data-driven framework for optical-spot disturbance modeling and tuning of covariance matrices of Kalman filters. Our approach is based on covariance estimation, nonlinear optimization, and subspace identification methods. Also, we use spectral factorization methods to emulate optical-spot disturbances with a desired power spectral density in an optical laboratory environment. We test the effectiveness of the proposed approaches on an experimental setup consisting of a piezo tip-tilt mirror, piezo linear actuator, and a CMOS camera.
\end{abstract}


\section{Introduction}

Stabilization, disturbance rejection, and precise control of optical beams and optical spots are fundamental and ubiquitous problems that appear in a number of applications and optical systems. For example, these problems are crucial for the development of sensing and control systems for space and ground optical telescopes~\cite{wie1993classical,hyde2004integrated,schwartz2018integrated,meza2005line,bely2003design,mosier2004role,acton2012wavefront,acton2018wavefront,redding1998wavefront,li2020theoretical,preda2018robust,eaton1993orbit}, free-space optical communication terminals~\cite{liang2021adaptive,kaushal2016optical}, turbulence compensation systems~\cite{liang2021adaptive}, precise laser beam steering and jitter compensation systems~\cite{tsuchiya2015receding,yoon2011laser,beerer2013practical, ahn2013evaluation,yoon2011laser,zhou2008design,skormin1995adaptive,watkins2007use}, and for other optical instruments and devices~\cite{kong2018vibration,li2018influence,chen2019simulating,chen2010modeling,patterson2015control,csencsics2017system}.

Optical spot and wave-front disturbances can originate from various sources and phenomena. In the case of ground telescopes, beside wavefront disturbances originating from the atmosphere, disturbances can originate from the wind and internal sources~\cite{bely2003design}. Internal disturbance sources are torque ripples in drive motors, friction in telescope axes, and the movement of mechanical components in instruments. According to~\cite{bely2003design}, the power spectrum of wind disturbances contains significant energy in the lower frequency range (from 0.1 to 1 Hz). Due to the fact that resonance frequencies of the structure and active mirror systems are either in this range or close to this range, the effect of wind disturbances can be significant. In the case of spacecraft or satellites carrying optical instruments, terminals, and sensors, the disturbances can originate from structural vibrations, actuator movement, and reaction wheel assemblies. These disturbances cause \textit{optical jitter}. According to~\cite{hyde2004integrated}, the term optical jitter is used to denote both stochastic image motion and observed dynamic wave-front errors. Control of jitter in optical instruments of space telescopes and satellites is a very challenging task. To illustrate these challenges, according to the authors of~\cite{hyde2004integrated}, the exposure time for James Webb Space Telescope can be up to 10,000 seconds. During this time interval, the tolerances for the wave-front error and line of sight motion are 14 nm and 4 mas, respectively. On the other hand, the target exposure time for the future Roman Space Telescope can be up to 30 hours, with a tolerance for the line of sight motion of 0.5 mas~\cite{mennesson2022roman}. In this manuscript, we are mainly concerned with the data-driven estimation of optical jitter coming from non-atmospheric sources. However, the methods proposed in this manuscript can be generalized to the case of atmospheric turbulence.  

Stringent requirements for rejecting optical disturbances and optical jitter call for the development of advanced model-based control and estimation  algorithms~\cite{massioni2015approximation,tesch2013receding,haber2022dual,hinnen2008data,polo2013linear,wie1993classical,Haber:13,haber2016framework,tsuchiya2015receding,beerer2013practical,Polo:12}. To develop such algorithms, it is first necessary to develop models and algorithms capable of optimally estimating, tracking, and predicting the spot position one or several samples in the future. The main challenge in developing such models and algorithms comes from the fact that the disturbance dynamics is often a combination of deterministic (periodic) and stochastic components~\cite{liu2008reaction}. Widely used methods for estimating and predicting state trajectories and disturbances of dynamical systems in mixed deterministic-stochastic environments are different versions of the Kalman filter~\cite{simon2006optimal}. Different versions and extensions of the Kalman filter are the backbones of many model-based algorithms, such as for example Linear Quadratic Gaussian (LQG) regulator~\cite{lewis2012optimal}.

When designing the Kalman filter, we are faced with at least two challenges. The first challenge is to develop a sufficiently accurate model of the system. In many cases, the first principle modeling approach might not produce sufficiently accurate models due to a lack of knowledge of the numerical values of model parameters or a complete lack of knowledge of the structure of equations describing the disturbance dynamics. One indirect approach for dealing with this problem is to develop Kalman filter models that approximate the disturbance dynamics as Newtonian systems. For example, such models assume that the velocity ($\boldsymbol{\alpha}-\boldsymbol{\beta}$ filter) or acceleration ($\boldsymbol{\alpha}-\boldsymbol{\beta}-\boldsymbol{\gamma}$ filter) are constant over a short time period~\cite{simon2006optimal}. However, these assumptions often lead to imprecise models, since the underlying assumptions of such models are often not met in practice. These Kalman filter models can be improved by the proper selection of the covariance matrices. Another approach for dealing with this problem is to employ data-driven techniques capable of updating the existing first-principle models or completely estimating the disturbance models from collected experimental data. System identification methods are often used for this purpose~\cite{verhaegen2007filtering}. However, it is challenging to apply the classical system identification methods to disturbance modeling since inputs to disturbance models are often unknown or they are partly or completely unpredictable.

The second challenge is that the covariance matrices of disturbance and measurement noise models are often not known \textit{a priori} or they often change with operating conditions. Since the knowledge of covariance matrices is directly used during the design of Kalman filters, this implies that imprecise knowledge of covariance matrices can significantly limit the performance of synthesized Kalman filters.

Motivated by these challenges, in this manuscript we propose and experimentally verify a unified data-driven framework and tools for the estimation, tracking, and prediction of optical spots by using data-driven Kalman filters. We address the first challenge by adapting, tuning, and experimentally testing the subspace system identification algorithm~\cite{haber2019subspace,haber2020modeling,haber2022subspace} for estimating the Kalman filter models of stochastic disturbances. We address the second challenge by developing a method for estimating the covariance matrices of Kalman filter models. Our covariance estimation approach is partly inspired by the autocovariance least-squares method~\cite{odelson2006new,rajamani2009estimation,kailath2000linear} and nonlinear optimization methods. Besides this, we demonstrate that the spectral factorization methods~\cite{anderson2012optimal,stoica2005spectral,kailath2000linear}, originally developed in control theory and signal processing communities, can be effective methods for emulating and experimentally producing mechanically-induced jitter disturbances with a desired spectrum in the laboratory environment. We test the effectiveness of the proposed approaches on an experimental setup consisting of a piezo tip-tilt mirror, piezo linear actuator, and a CMOS camera.

Subspace identification methods~\cite{verhaegen2007filtering} have been applied to the problem of estimating dynamical models of deformable mirrors used in adaptive optics systems, see for example~\cite{chiuso2009dynamic,Song2011paper,haber2020modelingHaberVerhaegen} and follow-up works. Furthermore, these methods have been applied to the problem of estimating thermally induced wavefront aberrations in ~\cite{haber2022subspace,haber2013identification,haber2013predictive,haber2020modeling,haber2022subspace}. The viability of using subspace identification methods for estimating atmospheric turbulence is analyzed in~\cite{hinnen2008data,hinnen2007exploiting}. To the best of our knowledge, little attention has been dedicated to investigating the viability of using subspace identification methods for building disturbance models of optical jitter coming from non-atmospheric sources.

This manuscript is organized as follows. In Section~\ref{section2}, we describe the experimental setup and explain the spectral factorization approach for producing the disturbances in the laboratory environment. In Section~\ref{section3}, we present the method for estimating the covariance matrices of Kalman filters. In Section~\ref{section4}, we present the subspace identification method for estimating Kalman filter models directly from the observed experimental data. In Section~\ref{section5}, we present conclusions and briefly discuss future work.

\section{Experimental Setup and Disturbance Emulation Using Spectral Factorization Approach}
\label{section2}

In this section, we describe the experimental setup that is used to verify the presented methods. We also explain and adapt the spectral decomposition method~\cite{anderson2012optimal,stoica2005spectral,kailath2000linear} for emulating the desired disturbance spectrum in the laboratory environment. 

\subsection{Experimental Setup}

The experimental setup is shown in Fig.~\ref{fig:Graph1} below. 

\begin{figure}[H]
	\centering 
	\includegraphics[scale=0.13,trim=0mm 0mm 0mm 0mm ,clip=true]{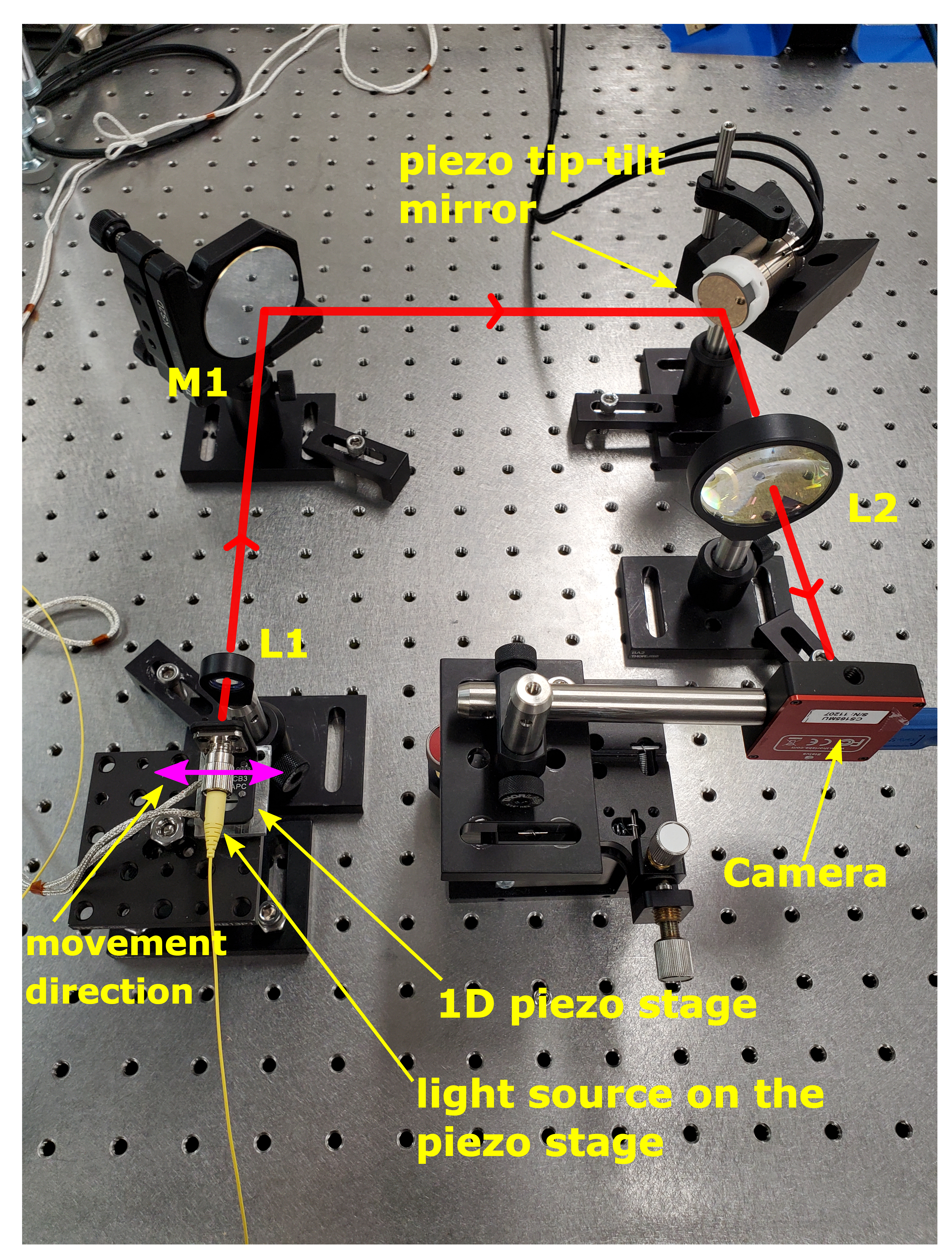}
	\caption{The experimental setup used to verify the presented methods.}
	\label{fig:Graph1}
\end{figure} 

In the sequel, we describe the main components of the constructed experimental setup. The red light fiber source is attached to an L-bracket that is attached to a piezo stage. We use the Micronix linear piezo stage. The product number is PP-12. The movement range of the stage is from $0$ to $4$ [mm] and the maximal speed is $3$ [mm/s]. The stage is equipped with an encoder with a resolution of $20$ [nm]. The piezo stage is controlled by the MMC-10 piezo motor controller. We tuned a PID controller and developed a MATLAB interface for controlling the piezo stage. By moving the piezo stage with the fiber source attached to it, we introduce horizontal optical spot disturbances (jitter) in the system. The lens L1 with a focal length of $25$ [mm] is used to collimate the beam. The beam is reflected from the mirror M1 and the piezo tip-tilt mirror. The piezo tip-tilt mirror is composed of a mirror attached to a tip-tilt piezo stage. We use the nPoint tip-tilt piezo stage, with the following specifications. The model number is RXY3-276. The travel range is +/- $1.5$ [mrad]. The resolution is $0.05$ [$\mu$rad] and the resonance frequency of the stage without the mirror is $2400$ [Hz]. To control the mirror, we use the LC.402 controller. In this paper, the tip-tilt piezo mirror is also used to introduce disturbance. This active mirror can introduce both horizontal and vertical disturbances in the camera detector plane, as well as a combination of such disturbances. In the follow-up publication, this mirror will be used as the correction element. The lens L2 is used to focus the beam onto a CMOS camera that is used as a detector. We use a Thorlabs CMOS camera with the product number CS165MU.

We use both the linear stage and the piezo tip-tilt mirror to introduce disturbances that are detected by the camera as stochastic spot movement. Our goal is to produce disturbances whose spectrum matches the spectrum of optical jitter movement that can be observed in an optical instrument mounted on a spacecraft, satellite, or on a vibrating platform. For that purpose, we need to use the spectral factorization method that is explained in the sequel.

\subsection{Disturbance emulation by using the spectral factorization approach}

The effect of disturbances on deployed optical systems can be directly measured by extracting time series data from the observed spot positions. Then, the statistical properties of the observed time series can be estimated. Also, disturbance information can be extracted by directly measuring the mechanical vibrations of the structure and system components. During the process of developing new and tuning the existing disturbance mitigation methods, it is necessary to reproduce such disturbances in a laboratory environment~\cite{liu2008reaction,bely2003design}. The main challenge is how to create disturbances in a laboratory environment whose statistical properties match the statistical properties of the disturbances affecting the deployed optical systems. Here, we address this important problem.

The statistical properties of disturbances are often described by power spectral density matrices. A power spectral density matrix can be estimated on the basis of the observed disturbance time series by using the methods summarized in~\cite{stoica2005spectral}. Let $S_{d}(z)$ be a power spectral density matrix that represents the spectral properties of the observed disturbances. Here, $z$ is a complex variable coming from the Z transform~\cite{kailath2000linear}. Examples of power spectral densities of real-life ground and space telescopes can be found in Chapter 7 of ~\cite{bely2003design} and in~\cite{liu2008reaction,mitani2014precision,yoshida2013spacecraft,dennehy2018spacecraft,pedreiro2003spacecraft,pedreiro2000next}. Having an estimate of $S_{d}(z)$, our main goal is to emulate such a disturbance spectrum in the laboratory environment. Our idea for solving this problem is to use and adapt the spectral factorization method~\cite{anderson2012optimal,stoica2005spectral,kailath2000linear} originally developed in control theory and signal processing communities.

\textbf{Spectral Decomposition Method.} The goal of the spectral decomposition method is to find a transfer function of the linear time-invariant stable system (filter) such that when a white noise sequence is applied to such a system, the power spectral density matrix of the output matches $S_{d}(z)$. From the mathematical point of view, we want to determine the transfer function matrix $H(z)$ of the system, such that 

\begin{align}
S_{d}(z)=H(z)S_{v}H^{T}(z^{-1})
\label{spectralFactorization}
\end{align}
where $S_{v}$ is the covariance matrix of the white noise (diagonal matrix, with the diagonal entries equal to the variance of the white noise inputs). That is, we want to decompose $S_{d}(z)$ in the form given by \eqref{spectralFactorization}.

Once we determine $H(z)$, we can apply a white noise sequence $\mathbf{v}_{k}$ to such system to compute the filtered output $\mathbf{d}_{k}$
\begin{align}
\mathbf{d}_{k}= H(q)\mathbf{v}_{k},
\label{simulatedOutput}
\end{align}
where $q$ is a shift delay operator~\cite{ljung1998system}, $H(q)$ is obtained by substituting $z$ by $q$ in $H(z)$, $\mathbf{d}_{k}$ is the output of the system, and $k$ is a discrete-time instant. Here, it should be emphasized that the power spectral density matrix of the filter $\mathbf{d}_{k}$ matches $S_{d}(z)$. 

This implies that under the assumption that the dynamics of the disturbance-generating element (such as a fast steering mirror or a piezo stage) can be neglected, the scaled version of the signal $\mathbf{d}_{k}$ can be used as a reference signal to the disturbance generating element. In this way, we can produce optical jitter disturbances whose spectrum matches $S_{d}(z)$ in the laboratory environment.The scaling factor of $\mathbf{d}_{k}$ can easily be determined.

The dynamics of the disturbance-generating element can be neglected if the sampling period of the sensor in the focal plane is larger than the settling time of the actuator. This is true in our case, since we are using a camera as a detector and its sampling frequency is typically below $100$ [Hz], and the settling time of the disturbance-generating elements is in the several millisecond range.

However, in the case of a fast spot detector or camera, the dynamics of the disturbance-generating element cannot be neglected. In this case, the output signal detected in the detector plane can be modeled as follows $\mathbf{d}_{k}=H_{a}(q) H(q)\mathbf{v}_{k}$, where $H_{a}(q)$ is the transfer function (dynamical model) of the actuator. Consequently, the signal $\mathbf{d}_{k}$ will not have the desired spectrum. We can deal with this problem by either modeling the actuator dynamics $H_{a}(q)$ by using first-principle approaches, or by estimating the actuator dynamics by using system identification methods~\cite{verhaegen2007filtering}. Then, we can invert this transfer function and multiply the reference signal by the inverse transfer function to obtain the output signal in the detector plane with the desired power spectral density matrix. 

The procedure for emulating the disturbance in the laboratory environment is summarized below.
\\
\textbf{Step 1: Obtain an estimate of the power spectral density matrix.} On the basis of the measurement of the disturbance affecting the deployed system or on the basis of the simulated model of the real system, obtain an estimate of the power spectral density matrix $S_{d}(z)$. 
\\
\textbf{Step 2: Perform spectral factorization.} Perform spectral factorization \eqref{spectralFactorization} to obtain the transfer function of the filter $H(z)$.
\\
\textbf{Step 3: Generate a reference signal for the disturbance-generating element.} Apply the white noise signal with the covariance matrix $S_{v}$ to the transfer function $H(z)$ to generate the output that is used as the reference signal for the disturbance-generating element. If necessary, scale or additionally modify such a signal to eliminate the effect of the actuator dynamics.

In this manuscript, we use an artificially constructed transfer function to validate the above-presented approach for emulating disturbances. This transfer function in the Laplace domain has the following form
\begin{align}
W& =W_{1}\cdot W_{2} \label{transferFunction}  \\
W_{1}& =\frac{10(s+\omega_{n1}^2)}{s^2+2\zeta_{1} \omega_{n1}s + \omega_{n1}^{2}},\;\; 
W_{2}=\frac{10(s+\omega_{n2}^2)}{s^2+2\zeta_{2} \omega_{n2}s + \omega_{n2}^{2}}, \notag
\end{align} 
where the parameters are given by $\omega_{n1}=2\pi\cdot 2 $, $\zeta_{1}=0.05$, $\omega_{n2}=2\pi\cdot 10$, and $\zeta_{2}=0.05$. The parameters $\omega_{ni}$, $i=1,2$, are called the natural undamped frequencies. On the other hand, the parameters $\zeta_{i}$, $i=1,2$, are called the damping ratios. Figure 2(a) shows the Bode magnitude and the phase plots of the transfer function \eqref{transferFunction}. The transfer function~\eqref{transferFunction} can model the disturbances produced by an elastic structure, external wind disturbances, or internal disturbance sources.

Our next goal is to construct the filter $H(z)$ that is obtained by decomposing a spectral density produced by \eqref{transferFunction}. There are at least two approaches to perform this decomposition. The first approach is to discretize~\eqref{transferFunction}, and then simulate this transfer function by applying the white noise input. Then, we can estimate the power spectral density of the output sequence. From the estimated power-spectral density, we can compute the function $H(z)$ by using the MATLAB function spectralfact().

The second approach that we pursue due to its simplicity, is described in the sequel. First, by using a discretization time step of $h=0.025$ seconds and a zero-order hold method~\cite{nise2020control}, we discretize the transfer function \eqref{transferFunction}. Let the discretized transfer function be denoted by $W_{d}(z)$. Then, we compute $S_{d1}=W_{d}(z)W_{d}(1/z)$. By using such $S_{d1}$, we compute \eqref{spectralFactorization}, by using the MATLAB function spectralfact(). This produces the following filter $H(z)$ and variance $S_{v}$

\begin{align}
H(z)=\frac{z^4 - 0.05229 z^3 - 0.3277 z^2 - 0.06164 z - 3.443 \cdot 10^{-10}}{z^4 - 1.876 z^3 + 1.831 z^2 - 1.604 z + 0.8282}, \; S_{v}=0.103. 
\label{computedFilterHSd}
\end{align}

\begin{figure}[H]
	\centering 
	\includegraphics[scale=0.8,trim=0mm 0mm 0mm 0mm ,clip=true]{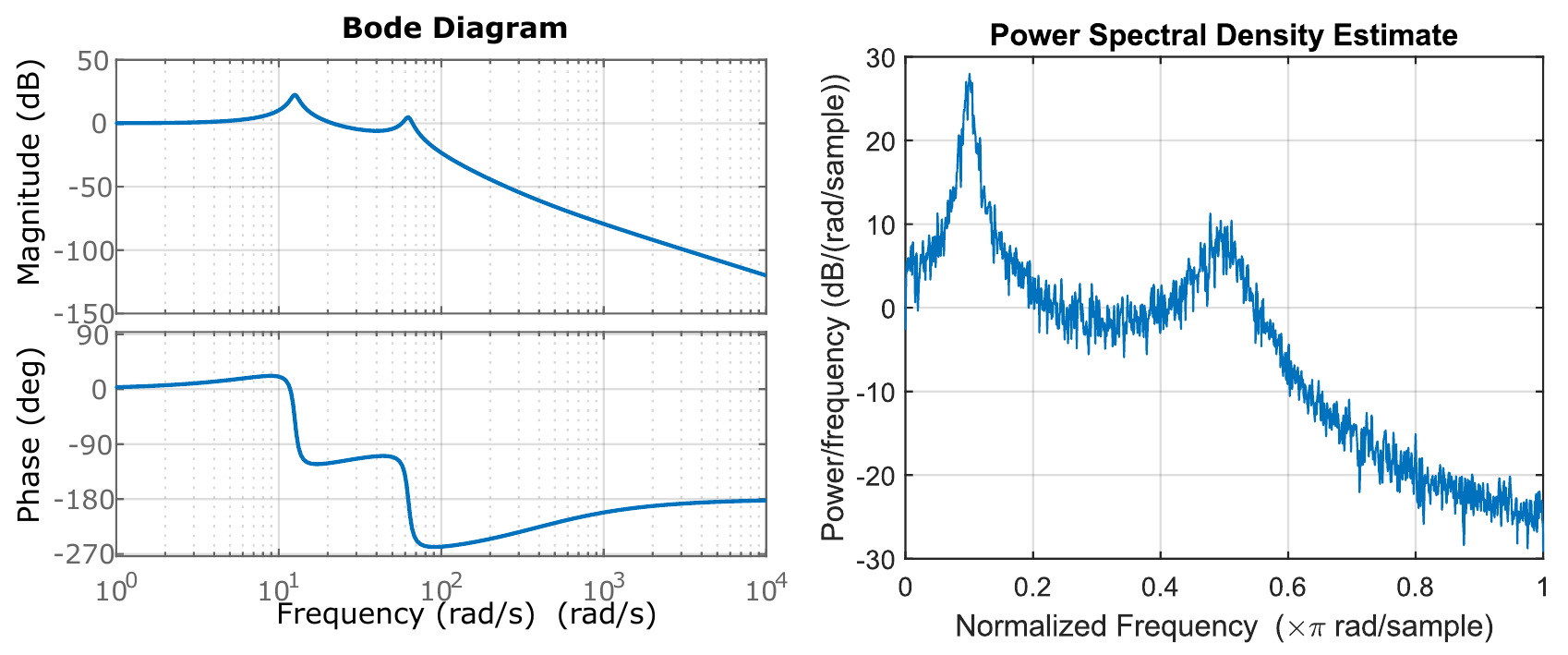}
	\caption{(a) The Bode diagram of the transfer function~\eqref{transferFunction}. (b) The power spectral density of the output $\mathbf{d}_{k}$ of the filter $H(z)$ defined in \eqref{computedFilterHSd} when the white noise sequence is applied as an input.}
	\label{fig:Graph2}
\end{figure}

Next, we simulate the computed filter \eqref{computedFilterHSd} with the white noise sequence $\mathbf{v}_{k}$ applied as the input, to obtain the sequence $\mathbf{d}_{k}$ that is used as the reference signal for the linear piezo actuator and the piezo tip-tilt mirror (disturbance generating elements). Figure~\ref{fig:Graph2}(b) shows the power spectral density of $\mathbf{d}_{k}$.

\section{Data Driven Tuning of Tracking Kalman Filter}
\label{section3}

In this section, we first present a data-driven approach for estimating the covariance matrices that are necessary for designing and tuning the Kalman filter. Then, we briefly summarize the approximate Kalman filter models for tracking the optical spot position. In the final part of this section, we experimentally verify the developed covariance estimation approach. The approach developed in this section is partly inspired by the autocovariance least-squares method~\cite{odelson2006new} and nonlinear optimization methods.

\subsection{Method for estimating the covariance matrices of the Kalman filter}
Consider the following state-space model 
\begin{align}
\mathbf{x}_{k+1}=A\mathbf{x}_{k}+G\mathbf{w}_{k}, \label{stateSpaceModel1} \\
\mathbf{y}_{k}=C\mathbf{x}_{k} + \mathbf{v}_{k},  \label{stateSpaceModel2} 
\end{align}

where $k$ is a discrete-time instant, $A\in \mathbb{R}^{n\times n}$, $G\in \mathbb{R}^{n\times s}$, and $C\in \mathbb{R}^{r\times n}$ are the system matrices, $\mathbf{x}_{k}\in \mathbb{R}^{n}$ is the state vector, $\mathbf{w}_{k}\in \mathbb{R}^{s}$ is the process disturbance (process noise) vector, $\mathbf{v}_{k}\in \mathbb{R}^{r}$ is the measurement noise vector, and $\mathbf{y}_{k}\in \mathbb{R}^{r}$ is the observed output vector. 

The covariance matrix of $\mathbf{w}_{k}$ is denoted by $Q\in \mathbb{R}^{s\times s}$. The covariance matrix of $\mathbf{v}_{k}$ is denoted by $R\in \mathbb{R}^{r\times r}$. We assume that $\mathbf{v}_{k}$ and $\mathbf{w}_{k}$ are uncorrelated. The covariance matrices $Q$ and $R$ are important since they are directly used to design the Kalman filter for the system given by the equations  \eqref{stateSpaceModel1} and \eqref{stateSpaceModel2}. However, these matrices are usually not known \textit{a priori}. Our goal is to develop a method for estimating these covariance matrices. In the case of separately tracking the $x$ and $y$ projections of the optical spot center, the vector $\mathbf{y}_{k}$ contains only a single entry that is equal to the observed $x$ or $y$ projections at the discrete time instant. That is, for every projection, we design a separate Kalman filter.  However, in the case of the system identification method that is developed in Section~\ref{section4}, the output vector $\mathbf{y}_{k}$ is two-dimensional and contains both $x$ and $y$ projections at the discrete-time instant $k$.

For the system given by equations \eqref{stateSpaceModel1}-\eqref{stateSpaceModel2}, we can design a (suboptimal) observer having the following form

\begin{align}
\hat{\mathbf{x}}_{k+1|k} & =A\hat{\mathbf{x}}_{k|k}, \label{subOptimalObserver1} \\
\hat{\mathbf{x}}_{k|k} & =\hat{\mathbf{x}}_{k|k-1} + L \underbrace{(\mathbf{y}_{k}-C\hat{\mathbf{x}}_{k|k-1})}_{\mathbf{e}_{k}}, \label{subOptimalObserver2}
\end{align}

where $L\in \mathbb{R}^{n\times r}$ is the observer gain. The vector $\hat{\mathbf{x}}_{k|k}$ is the state estimate that takes into account previous state estimates and observed outputs up to the time instant $k$. This vector is also called the \textit{a posteriori} estimate. The vector $\hat{\mathbf{x}}_{k+1|k}$ is the state estimate at the time instant $k+1$ computed on the basis of the estimates and observed outputs up to the time instant $k$. This vector is also called the \textit{a priori} state estimate. The quantity $\mathbf{e}_{k}\in \mathbb{R}^{r}$
\begin{align}
\mathbf{e}_{k}=\mathbf{y}_{k}-C\hat{\mathbf{x}}_{k|k-1},
\label{innovationVector}
\end{align}
that appears in \eqref{subOptimalObserver2}, is called the \textit{innovation vector}. After substituting (\ref{subOptimalObserver2}) in (\ref{subOptimalObserver1}), we obtain

\begin{align}
\hat{\mathbf{x}}_{k+1|k} & =A\hat{\mathbf{x}}_{k|k-1} + AL \mathbf{e}_{k} \label{substitutedState}.
\end{align}

The estimation error is defined as follows

\begin{align}
\boldsymbol{\varepsilon}_{k}=\mathbf{x}_{k}-\hat{\mathbf{x}}_{k|k-1}.
\label{estimationError}
\end{align}

By propagating \eqref{estimationError} one time step, and by substituting \eqref{stateSpaceModel1}, \eqref{stateSpaceModel2}, and \eqref{substitutedState} in the resulting equation, we obtain 

\begin{align}
\boldsymbol{\varepsilon}_{k+1} & = \bar{A} \boldsymbol{\varepsilon}_{k} +\bar{G} \bar{\mathbf{w}}_{k}, \label{errorDynamics1} \\
\mathbf{e}_{k} & =C \boldsymbol{\varepsilon}_{k}+\mathbf{v}_{k},
\end{align}
where
\begin{align}
\bar{A}=A-ALC,\;\; \bar{G}=\begin{bmatrix} G & -AL \end{bmatrix},\;\; \bar{\mathbf{w}}_{k}=\begin{bmatrix} \mathbf{w}_{k}  \\ \mathbf{v}_{k}  \end{bmatrix}.
\label{explanation}
\end{align}

Next, we introduce the autocorrelation matrices

\begin{align}
\mathcal{A}_{j}=E\big[\mathbf{e}_{k} \mathbf{e}_{k+j}^{T} \big], \;\; j=0,1,\ldots, N_{A},
\label{autocorrelations}
\end{align}
where $\mathcal{A}_{j}\in \mathbb{R}^{r\times r}$, and $N_{A}$ is the total number of autocorrelation matrices. It can be shown that under a steady-state assumption~\cite{odelson2006new}, we have

\begin{align}
\mathcal{A}_{0}& =CPC^{T}+R, \label{autocorrelation1} \\
\mathcal{A}_{j}& =C\bar{A}^{j}PC^{T}-C\bar{A}^{j-1}ALR, \;\; j=1,2,\ldots,N_{A},\label{autocorrelation2}
\end{align}

where $P$ is the steady-state covariance matrix of the estimation error $\boldsymbol{\varepsilon}_{k}$. This covariance matrix satisfies the following equation 

\begin{align}
P=\bar{A}P\bar{A}^{T}+GQG^{T}+ALRL^{T}A^{T}.
\label{lyapEquation1}
\end{align}

On the basis of \eqref{autocorrelations}, we can estimate the correlation matrices $\mathcal{A}_{j}$ as follows

\begin{align}
\hat{\mathcal{A}}_{j}=\frac{1}{N_{A}}\sum_{i=0}^{N_{A}-j} \mathbf{e}_{i} \mathbf{e}_{i+j}^{T},
\label{covarianceMatricesEstimation}
\end{align}
where $\hat{\mathcal{A}}_{j}$ denotes an estimate of $\mathcal{A}_{j}$. 

Our idea for estimating the covariance matrices is summarized in the sequel. First, we substitute the "true" correlation matrices $\mathcal{A}_{j}$ by their estimates $\hat{\mathcal{A}_{j}}$ in \eqref{autocorrelation1} and \eqref{autocorrelation2}. Then, \eqref{autocorrelation1}, \eqref{autocorrelation2}, and \eqref{lyapEquation1} represent a system of equations with the unknowns $P$,$Q$, and $R$. The first step in solving such a system is to use \eqref{lyapEquation1} to express $P$ as the function of $Q$ and $R$, and then to substitute $P$ in \eqref{autocorrelation1} and \eqref{autocorrelation2}. Then, the resulting equation is solved by formulating a nonlinear optimization problem with unknowns $Q$ and $R$.

Let $\otimes$ denote the Kronecker vector product and let $\text{vec}(\cdot)$ denote the vectorization operator~\cite{verhaegen2007filtering}. If $M\in \mathbb{R}^{n\times m}$ is a matrix, then $\text{vec}(M)\in \mathbb{R}^{n\cdot m}$ is an $n\cdot m$ vector obtained by stacking the entries of the matrix $M$ column wise on top of each other and starting from the first column of $M$. If $Z_{1},Z_{2}$, and $Z_{3}$ are arbitrary matrices, then we have 
\begin{align}
\text{vec}(Z_{1}Z_{2}Z_{3})=\big(Z_{3}^{T}\otimes Z_{1}\big) \text{vec}(Z_{2}).
\label{operator1}
\end{align}

By applying the vectorization operator to \eqref{lyapEquation1}, we obtain 

\begin{align}
\text{vec}(P)=\big(I-\bar{A}_{1} \big)^{-1}G_{1}\text{vec}(Q)+\big(I-\bar{A}_{1} \big)^{-1}\bar{A}_{2}\text{vec}(R),
\label{vectorizationOperator1} \\
\bar{A}_{1}=\bar{A}\otimes\bar{A}, \;\; G_{1}=G\otimes G, \;\; \bar{A}_{2}=AL\otimes AL. \label{vectorizationOperator1Explanation}
\end{align}

More details and examples of using the vectorization and the Kronecker operators can be found in~\cite{haber2014sparseLyapunov,haber2018sparsity}. Then, by applying the vectorization operator to \eqref{autocorrelation1} and \eqref{autocorrelation2}, and by substituting $\text{vec}(P)$ from \eqref{vectorizationOperator1} in the resulting equations, we obtain 

\begin{align}
\hat{\mathbf{a}}_{0}=\text{vec}(\hat{\mathcal{A}}_{0})=C_{1}\big(I-\bar{A}_{1} \big)^{-1}G_{1}\text{vec}(Q)+\Big(I+C_{1}\big(I-\bar{A}_{1} \big)^{-1}\bar{A}_{2} \Big)\text{vec}(R), \label{autoExpressed1} \\
C_{1}=C\otimes C, \label{autoExpressed1Explained}
\end{align}
and 
\begin{align}
\hat{\mathbf{a}}_{j}=\text{vec}(\hat{\mathcal{A}}_{j})=\bar{A}_{3,j} \big(I-\bar{A}_{1} \big)^{-1}G_{1}\text{vec}(Q)+\Big(\bar{A}_{3,j}\big(I-\bar{A}_{1} \big)^{-1}\bar{A}_{2}-\bar{A}_{4,j-1} \Big)\text{vec}(R)
\label{autoExpressed2} \\
\bar{A}_{3,j}=\big(C\otimes C\bar{A}^{j} \big),\;\; \bar{A}_{4,j-1}=I\otimes \big(C\bar{A}^{j-1}AL \big),\;j=1,2,\ldots, N_{A}.
\label{autoExpressed2Explained}
\end{align}
The set of equations \eqref{autoExpressed1} and \eqref{autoExpressed2}, can be written compactly
\begin{align}
\hat{\mathbf{a}} & =H \mathbf{w}, \label{compactEquation} \\
\hat{\mathbf{a}}& =\begin{bmatrix}\hat{\mathbf{a}}_{0} \\ \hat{\mathbf{a}}_{1} \\ \vdots \\ \hat{\mathbf{a}}_{N_{A}} \end{bmatrix}, H=\begin{bmatrix} C_{1}\big(I-\bar{A}_{1} \big)^{-1}G_{1} & I+C_{1}\big(I-\bar{A}_{1} \big)^{-1}\bar{A}_{2} \\ \bar{A}_{3,1} \big(I-\bar{A}_{1} \big)^{-1}G_{1} &  \bar{A}_{3,1}\big(I-\bar{A}_{1} \big)^{-1}\bar{A}_{2}-\bar{A}_{4,0} \\ \vdots & \vdots \\ 
\bar{A}_{3,N_{A}} \big(I-\bar{A}_{1} \big)^{-1}G_{1} &  \bar{A}_{3,N_{A}}\big(I-\bar{A}_{1} \big)^{-1}\bar{A}_{2}-\bar{A}_{4,N_{A}-1}	\end{bmatrix}, \mathbf{w}=\begin{bmatrix}\text{vec}(Q) \\ \text{vec}(R) \end{bmatrix}.	\label{compactEquationExplanation}
\end{align}

We determine the covariance matrices whose entries are the entries of the vector $\mathbf{w}$ by solving the following optimization problem
\begin{align}
& \min_{\mathbf{w}} \left\|\hat{\mathbf{a}} - H \mathbf{w}  \right\|_{2}^{2},  \label{optimizationProblem1} \\
& \text{subject to:} \notag  \\
& Q=Q^{T}, R=R^{T}, \label{optimizationProblem12}  \\
& Q\ge 0, R \ge 0, \label{optimizationProblem13} \\
& \mathbf{b}_{1} \le  \text{vec}(Q)\le \mathbf{b}_{2},\mathbf{b}_{3} \le  \text{vec}(R)\le \mathbf{b}_{4},   \label{optimizationProblem14}
\end{align} 
where $\mathbf{b}_{1}$ and $\mathbf{b}_{2}$ are the lower and upper bounds on the entries of the matrix $Q$, and $\mathbf{b}_{3}$ and $\mathbf{b}_{4}$ are the lower and upper bounds on the entries of the matrix $R$. The constraints in the above optimization problem are used to enforce positive semi-definiteness on the matrices $Q$ and $R$.

In the sequel, we explain how the above-stated optimization problem can be transformed into a form that can be numerically implemented and solved. The main challenge is how to transform the constraints $Q\ge 0$ and $R\ge 0$ into a numerically tractable form. In this manuscript, we decouple the tracking of optical spots in the x and y directions. Consequently, for each direction, the matrix $R$ is actually a scalar. Due to this, the positive semi-definite constraint $R\ge 0$, simply translates to the condition that the scalar $R$ is greater or equal to zero. 

On the other hand, in this manuscript, we consider three possible cases for the matrix $Q$, and we explain how to implement the constraint $Q\ge 0$.
\\\\
\textbf{Case 1: $Q\in \mathbb{R}^{1\times 1}$}. In this case, the disturbance vector $\mathbf{w}_{k}$ is one dimensional. Consequently, $Q$ is a scalar and the constraint $Q\ge 0$ is a scalar constraint that can be directly incorporated into the optimization solver. 
\\
\textbf{Case 2: $Q\in \mathbb{R}^{2\times 2}$}. In this case, the disturbance vector $\mathbf{w}_{k}$ is two-dimensional. It is well-known that a matrix is positive semi-definite if all the principal minors are greater than or equal to zero. This translates to the following conditions for the entries of the matrix $Q$:
\begin{align}
q_{11}\ge 0, \;\; q_{22} \ge 0 , \;\; q_{11}q_{22}-q_{12}q_{21}\ge 0,
\label{positiveSemidefinite1}
\end{align}
where $q_{11}$, $q_{12}$, $q_{21}$, and $q_{22}$ are the corresponding entries of the matrix $Q$.
\\
\textbf{Case 3: $Q\in \mathbb{R}^{3\times 3}$}. In this case, the disturbance vector $\mathbf{w}_{k}$ is three-dimensional. Similarly to the two-dimensional case, the positive semi-definite condition is that all principal minors should be greater than or equal to zero. This translates into the following conditions:
\begin{align}
& q_{11}\ge 0 , q_{22}\ge 0, q_{33}\ge 0, \label{positiveSemidefinite21} \\
& q_{22}q_{33}-q_{32}q_{23}\ge 0, q_{11}q_{33}-q_{31}q_{13} \ge 0, q_{11}q_{22}-q_{21}q_{12}\ge 0, \label{positiveSemidefinite22}\\
& \text{det}(Q)\ge 0, \label{positiveSemidefinite23}
\end{align}

where $q_{11},q_{12},\ldots, q_{33}$ are the corresponding entries of $Q$, and $\text{det}(Q)$ is the determinant of $Q$.

\subsubsection{Summary of the procedure for estimating the covariance matrices}
The procedure for estimating the covariance matrices is summarized below. 
\\
\\
\textbf{Step 1: Design the suboptimal observer gain $L$.} Use the pole placement method to compute the suboptimal observer gain $L$ of the observer \eqref{substitutedState}. This step can be performed in MATLAB by using the function place().
\\
\textbf{Step 2: Estimate the correlation values and solve the optimization problem.} Use the designed observer to compute the innovation sequence \eqref{innovationVector}. Estimate the autocorrelation matrices by using \eqref{covarianceMatricesEstimation}. Then, solve the optimization problem defined by the equations \eqref{optimizationProblem1} to \eqref{optimizationProblem14}. The optimization problem can be solved by using the MATLAB function fmincon(). In this step, the estimates of the matrices $Q$ and $R$ are computed.
\\
\textbf{Step 3: Compute the Kalman filter gain.} On the basis of the estimated matrices $Q$ and $R$, compute the Kalman filter gain. This step is performed by solving the Riccati equation for the computed $Q$ and $R$. The Riccati equation can be solved by using the MATLAB Riccati solver dare(). Let the solution of the Riccati equation be denoted by $P_{r}$. Then, the data-driven Kalman filter gain is obtained as follows
\begin{align}
L_{K}=P_{r}C^{T}\big(CPC^{T}+R\big)^{-1}.
\label{dataDrivenKalmanFilter}
\end{align} 

Once the data-driven Kalman filter gain is computed, we can substitute $L_{K}$ instead of the suboptimal gain $L$ in the observer equation \eqref{substitutedState}, to obtain an approximate Kalman filter. \textit{Here, it is important to emphasize that it is a good practice to iteratively perform steps 2 and 3 of the above-presented procedure.} Namely, once we compute $L_{K}$ in step 3, we can use this matrix $L_{k}$ instead of the suboptimal gain $L$ in step 2. In this way, we can get even better estimation results. We follow this strategy to iteratively improve the estimates of the matrices $Q$ and $R$.

\subsection{Tracking Kalman Filter}

In the previous subsection, we presented the procedure for estimating the covariance matrices. Here, we present equations describing a tracking Kalman filter model that is combined with the covariance estimation procedure. The presented Kalman filter is an adapted version of the three-dimensional $\alpha$-$\beta$-$\gamma$ filter~\cite{simon2006optimal}.

Let $h>0$ be a small discretization constant that is at the same the sampling period of the Kalman filter. The filter is defined by the following state-space matrices in \eqref{stateSpaceModel1} and \eqref{stateSpaceModel2}:

\begin{align}
A=\begin{bmatrix} 1& h & \frac{h^2}{2} \\ 0 & 1 & h \\ 0 & 0 & 1 \end{bmatrix},\; C=\begin{bmatrix} 1 & 0 & 0 \end{bmatrix}, \; G=\begin{bmatrix} \frac{h^{2}}{2}  \\ h \\ 1 \end{bmatrix}, Q=\sigma_{w}^{2}, R=\sigma_{v}^{2},
\label{thirdOrderKalmanFilter}
\end{align} 

where $\sigma_{w}$ and $\sigma_{v}$ are the standard deviations of the disturbance and measurement noise.
The first, second, and third state variables of this filter are position, velocity, and acceleration. This filter assumes that the noisy position is measured.  The filter matrix $A$ is obtained from simple kinematic equations of a particle.  Note that without the disturbance $\mathbf{w}_{k}$, this model assumes that the derivative of the acceleration is constant. However, this might not occur in practice. Consequently, the process disturbance noise is included to relax the model assumption. The structure of the matrix $G$ is a direct consequence of this assumption. 

The matrix $G$ can be integrated into the covariance matrix. Namely, it has been shown that from the Kalman filter perspective, the model \eqref{thirdOrderKalmanFilter} is equivalent to the following model

\begin{align}
\mathbf{x}_{k+1}=A\mathbf{x}_{k}+\tilde{\mathbf{w}}_{k}, \label{stateSpaceModel1Modified} \\
\mathbf{y}_{k}=C\mathbf{x}_{k} + \mathbf{v}_{k},  \label{stateSpaceModel2Modified} 
\end{align}

with the covariance matrix of $\tilde{\mathbf{w}}_{k}$ given by 

\begin{align}
\tilde{Q}=GQG^{T},	
\end{align}

where $Q$ is the covariance matrix of the original distrubance vector $\mathbf{w}_{k}$. By using this idea, we obtain the following covariance matrix

\begin{align}
\tilde{Q}=\begin{bmatrix}\frac{1}{4}h^4 & \frac{1}{2}h^3 &  \frac{1}{2}h^2 \\ \frac{1}{2}h^3 & h^2 & h \\ \frac{1}{2}h^2 & h & 1   \end{bmatrix}\sigma_{w}^{2},\; R=\sigma_{v}^2.
\label{covarianceMatrix23}
\end{align}

However, the type of the covariance matrix given by \eqref{covarianceMatrix23} is not favorable from the estimation perspective. The issue is that the dimension of the estimation problem is increased by having to estimate 9 entries of the matrix $\tilde{Q}$ compared to only a single entry of the matrix $Q$ that needs to be estimated in the case of the model \eqref{thirdOrderKalmanFilter}. Consequently, in this manuscript, we keep the original formulations of the Kalman filter and the covariance matrix given by \eqref{thirdOrderKalmanFilter}. That is, only the scalars $\sigma_{w}$ and $\sigma_{v}$ need to be estimated.

\subsection{Experimental Results}

We present the experimental results of using the developed covariance estimation approach and the tracking Kalman filter. 

Here, it is important to first test the developed approach on a deterministic signal applied to the disturbance generator. This is because the deterministic signal helps us to better understand and visualize the tracking properties of the developed data-driven Kalman filter. However, in the case of the linear piezo stage disturbance generator, despite the fact that we apply the deterministic signals as a reference input, the spot observed by the camera is still partly stochastic due to the fact that the linear piezo stage produces additional vibrations. In the next section, we test the identification algorithm on completely stochastic signals generated by the spectral factorization approach.

We apply a sinusoidal reference signal to the linear piezo stage. The reference signal is defined by 
\begin{align}
r(t)=0.1\sin(2t)+0.05\sin(6t)+2.
\label{referenceSignal}
\end{align}
The spot positions are observed by the camera. The mean sampling period of the camera is $0.0177$ seconds. This sampling period slightly varies due to the fact that it is challenging to precisely control the sampling frequency on a Windows computer with MATLAB. 

We use the well-known center of the mass algorithm to detect the spot center. We apply a combination of two $\alpha$-$\beta$-$\gamma$ filters and the developed covariance estimation method to the observed $x$ and $y$ projections of the center point. The matrices $Q$ and $R$ are scalars. In the tracking Kalman filter and for the estimation of the covariance matrices, we use the discretization constant of $h=0.0177$ and the number of autocorrelation coefficients of $N_{A}=200$. We start with a suboptimal observer designed by placing the eigenvalues in the set $\{0.3,0.4, 0.5\}$ (step 1 of the procedure summarized in Section 3.1.1). We iteratively perform steps 2 and 3 of the estimation procedure, where in every iteration we start with previously computed observer gains. We perform 10 iterations.

\begin{figure}[H]
	\centering 
	\includegraphics[scale=0.38,trim=0mm 0mm 0mm 0mm ,clip=true]{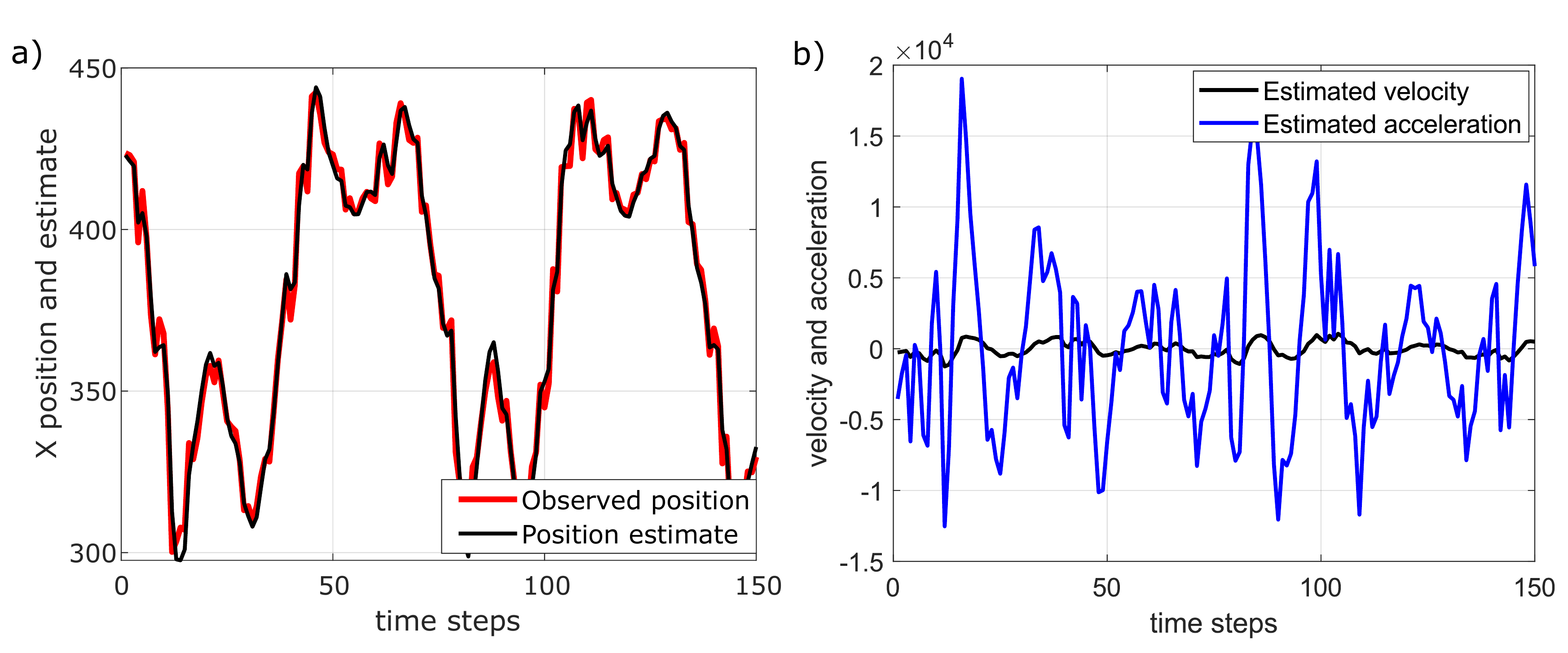}
	\caption{(a) Observed and estimated $x$ projection of the spot center point. (b) Estimated velocity and acceleration of the $x$ projection.}
	\label{fig:Graph3}
\end{figure}

\begin{figure}[H]
	\centering 
	\includegraphics[scale=0.4,trim=0mm 0mm 0mm 0mm ,clip=true]{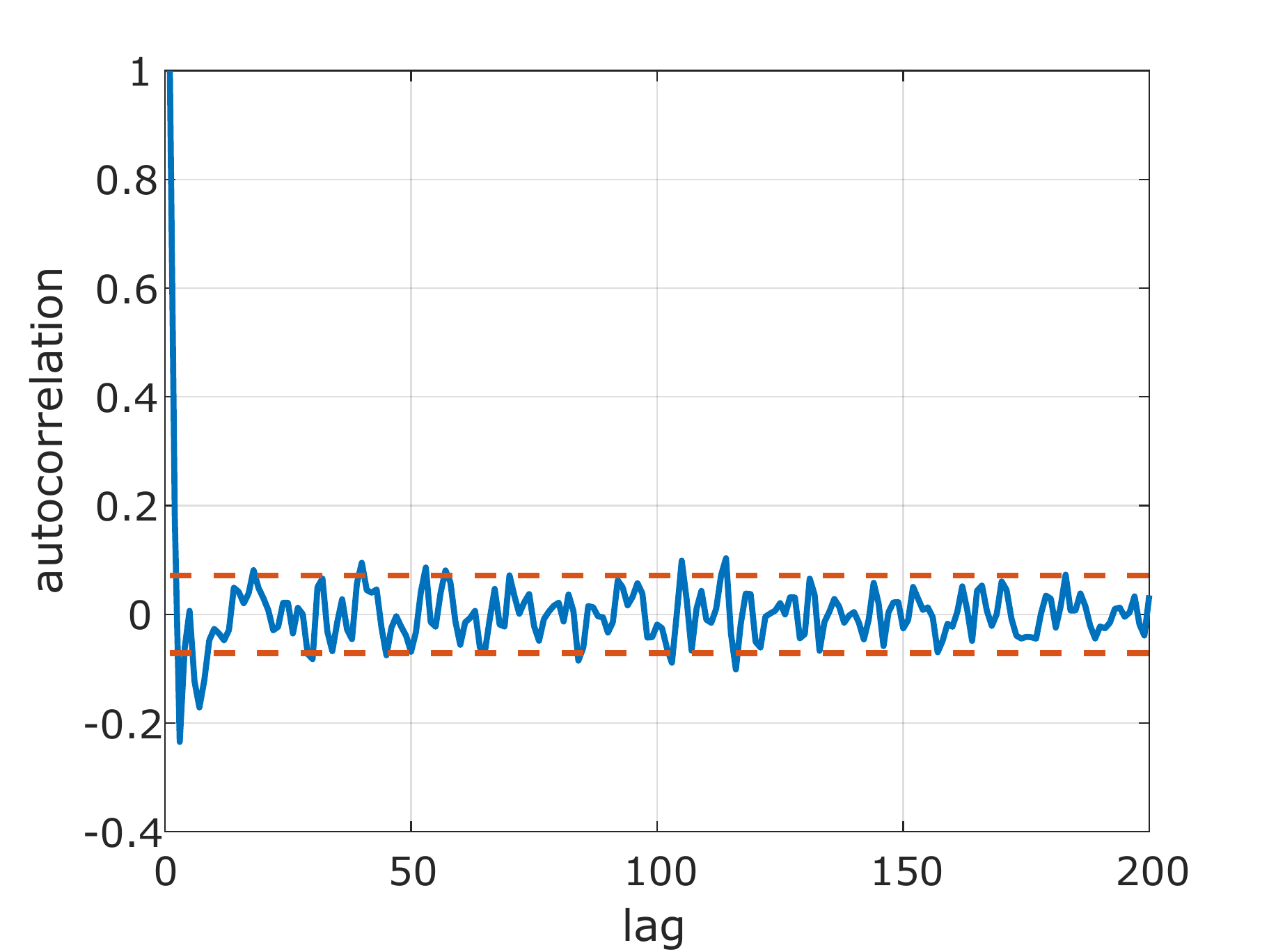}
	\caption{Autocorrelation coefficients of the innovation sequences. The red dashed lines are limits used for white-noise hypothesis testing.} 
	\label{fig:Graph4}
\end{figure}

Figure~\ref{fig:Graph3}(a) shows the comparison between the observed and estimated $x$ projections of the spot center. Figure~\ref{fig:Graph3}(b) shows the estimated velocity and acceleration of the spot center. Here it should be kept in mind that we only observe the position, and the velocity and acceleration are estimates that cannot be compared with the observed values. Figure~\ref{fig:Graph4} shows the autocorrelation coefficients of the innovation sequence. In an ideal case, the innovation sequence should be a white noise sequence. This means that the autocorrelation coefficients should be $1$ for the lag zero, and otherwise zero or very close to zero. This ideal case is achieved under the assumptions that (1) the covariance matrices are perfectly estimated, (2) the measurement noise and disturbances are white Gaussian noise sequences, and (3) the computed Kalman gain is optimal. Consequently, we can perform a white-noise hypothesis test on the innovation sequence in order to investigate the optimality of the Kalman filter and evaluate the performance of the covariance estimation procedure. The red dashed lines in Fig.~\ref{fig:Graph4} represent the limits of the autocorrelation coefficients. If 95~$\%$ of autocorrelation coefficients are inside of the region bounded by the red dashed lines, we can conclude that the innovation sequence is a white noise sequence. There is a total of 21 autocorrelation coefficients out of 200 coefficients that exceed the bounds. This is roughly 10~$\%$ of the coefficients. Consequently, we can conclude that the residual does not have a white noise property. However, this number is still significantly smaller than the number of autocorrelation coefficients of the initial suboptimal observer with the gain of $L$. This is a good indication that the proposed data-driven method actually works in practice and that it is a viable tool for tuning the Kalman filters.

\section{System identification of the Kalman Filter state-space models}
\label{section4}

In the previous section, we used the Kalman filters that are based on approximate first-principle models of the spot dynamics. The previously used models are based on the Newtonian assumption on spot dynamics. By tuning the covariance matrices, we can improve the performance of the filters and partly compensate for the model inaccuracies. However, this approach has limitations originating from the modeling assumptions. In this section, we use a completely different approach for building the models and deriving the Kalman filter state-space models. Instead of assuming the model parameters \textit{a priori}, we use a data-driven approach to estimate the system model and the (suboptimal) Kalman filter gain directly from the observed spot position time series. The approach used in this section relies upon the subspace identification approach~\cite{haber2014subspace,haber2019subspace,haber2019identification,haber2020modeling}.

In particular, we base our identification approach on a modified version of the subspace identification algorithm that is presented in~\cite{haber2020modelingHaberVerhaegen}. The main modification is that we eliminate the exogenous inputs from the subspace identification method and we use past outputs as inputs.

\subsection{Summary of the Subspace Identification Method}
To simplify the notation in \eqref{substitutedState}, we substitute $\hat{\mathbf{x}}_{k|k-1}$ by $\hat{\mathbf{x}}_{k}$. Then, from \eqref{substitutedState} we obtain the following model
\begin{align}
\hat{\mathbf{x}}_{k+1}& =\bar{A}\hat{\mathbf{x}}_{k}+\tilde{L}\mathbf{y}_{k}, \label{finalEquationKalman1} \\
\mathbf{y}_{k}& =C\hat{\mathbf{x}}_{k},
\label{finalEquationKalman2}
\end{align}
where $\bar{A}=A-ALC$ and $\tilde{L}=AL$. We assume that the output vector $\mathbf{y}_{k}$ is two-dimensional with the entries equal to $x$ and $y$ projections of the optical spot center.

\textit{\textbf{Subspace Identification Problem:} From the sequence $\{\mathbf{y}_{i}\}^{i=0,1,2,\ldots, N}$ of the observed projections of the spot center point, estimate the model order $n$, and the system matrices $\bar{A}$, $A$, $\tilde{L}$, and $C$ of the state-space model~\eqref{finalEquationKalman1} and \eqref{finalEquationKalman2}.}

To summarize the modified subspace identification algorithm, we need to introduce the following notation that is originally introduced in~\cite{haber2020modelingHaberVerhaegen}. Let $i_{1}$, $i_{2}$, and $l$ be three positive integers corresponding to discrete-time instants. We introduce the following notation:
\begin{align}
\mathbf{y}_{i_{1},i_{2}}=\begin{bmatrix} \mathbf{y}_{i_{1}} \\ \mathbf{y}_{i_{1}+1} \\ \vdots \\ \mathbf{y}_{i_{2}}  \end{bmatrix}, \;\;  Y_{i_{1},i_{2}}^{(l)}=\begin{bmatrix} \mathbf{y}_{i_{1},i_{2}} & \mathbf{y}_{i_{1}+1,i_{2}+1} & \ldots & \mathbf{y}_{i_{1}+l,i_{2}+l} \end{bmatrix}. \label{matrixDefinintionLifted}
\end{align}

The subspace identification method is presented below.
\\
\\
\textbf{Step 1: Estimation of the Markov matrices.} Choose the past window $p$ and the parameter $l$ as $l=N-p-1$, and estimate the matrix of Markov parameters $M_{p}$ as follows:
\begin{align}
\hat{M}_{p}= Y_{p,p}^{(l)}Y_{0,p-1}^{\dagger},
\label{estimateOfMarkovParameters}
\end{align}
where the symbol $\dagger$ denotes the matrix pseudo-inverse. The past window $p$ is selected on the basis of the Akaike Information Criterion (AIC), for more details see the experimental results below and \cite{haber2020modelingHaberVerhaegen}.
\\
\textbf{Step 2: Estimate the state sequence.} Select the future window parameter $f$, such that $f \le p$, and form the matrix $\hat{\mathcal{M}}$ from the estimated Markov parameters as follows:
\begin{align}
\hat{\mathcal{M}}=\begin{bmatrix} \hat{M}_{p} \\ 0_{r\times r} \;\;\;\;  \hat{M}_{p}(:,1:(p-1)r) \\ 0_{r\times 2r} \;\;\;\;  \hat{M}_{p}(:,1:(p-2)r) \\ \vdots  \\ 0_{r\times (f-1)r} \;\;\;\;  \hat{M}_{p}(:,1:(p-f+1)r) \end{bmatrix}.
\label{bigMatrixEstimate}
\end{align}
Then, compute the singular value decomposition~\cite{verhaegen2007filtering} of the matrix $\mathcal{D}$ 
\begin{align}
\mathcal{D}=\mathcal{U}\Sigma \mathcal{V}^{T},\;\; \mathcal{D}=\hat{\mathcal{M}}Y_{0,p-1}^{(l)}.
\label{singularValueDecomposition}
\end{align}
Select the state order $n$, and compute an estimate of the state sequence as follows 
\begin{align}
\hat{X}_{p,p}^{(l)}=\Sigma(1:n,1:n)^{1/2}\mathcal{V}(1:n,:).
\label{stateSequenceEstimate}
\end{align}
where the matrix $\hat{X}_{p,p}^{(l)}$ is defined in the similar manner to the matrix $Y_{i_{1},i_{2}}^{(l)}$ in \eqref{matrixDefinintionLifted} (the meaning the subscripts and superscripts is identical), with the difference that the output $\mathbf{y}_{k}$ is replaced by the estimated state $\hat{\mathbf{x}}_{k}$.
\\
\textbf{Step 3: Estimate the system matrices.} First, compute the following matrices
\begin{align}
\mathcal{X}_{1}=\begin{bmatrix} \hat{X}_{p,p}^{(l-1)} \\ Y_{p,p}^{(l-1)}  \end{bmatrix}, \; \mathcal{X}_{2}=\hat{X}_{p+1,p+1}^{(l-1)}\mathcal{X}_{1}^{\dagger}.
\label{systmeMatrices}
\end{align}
Then, estimate the system matrices as follows
\begin{align}
\hat{\bar{A}}=	\mathcal{X}_{2}(:,1:n),\;\; \hat{C}=Y_{p,p}^{(l)}\Big(\hat{X}_{p,p}^{(l)}\Big)^{\dagger },\;\; \hat{\tilde{L}}=\mathcal{X}_{2}(:,n+1:n+r),\; \hat{A}=\hat{\bar{A}}+\hat{\tilde{L}}\hat{C}.
\label{systemMatrices}
\end{align}

In step $2$, we need to estimate the state-order $n$ of the model. We estimate the state order on the basis of the singular value plot of the matrix $\mathcal{D}$. The number of most dominant singular values can be used as a good estimate of the state-order~\cite{verhaegen2007filtering}.

\subsection{Experimental results of applying the subspace identification algorithm}
Here, we present the experimental tests of the subspace identification algorithm. 

In the first case, we apply two inputs to the piezo tip-tilt mirror in Fig.~\ref{fig:Graph1}. These inputs are designed by using the spectral factorization method explained in Section~\ref{section2}. The power spectral density of these discrete-time signals is shown in Fig.~\ref{fig:Graph2}(b). We collect 4000 images by using the camera. By using the center of the mass algorithm, we extract $x$ and $y$ coordinates of spots. We split the collected data into two sequences. The first sequence of the length of $2000$ is used to identify the model. This sequence is called the identification data set. The second sequence of the length of $200$ is used to validate the model and to properly choose the model parameters. This sequence is called the validation sequence.

\begin{figure}[H]
	\centering 
	\includegraphics[scale=0.3,trim=0mm 0mm 0mm 0mm ,clip=true]{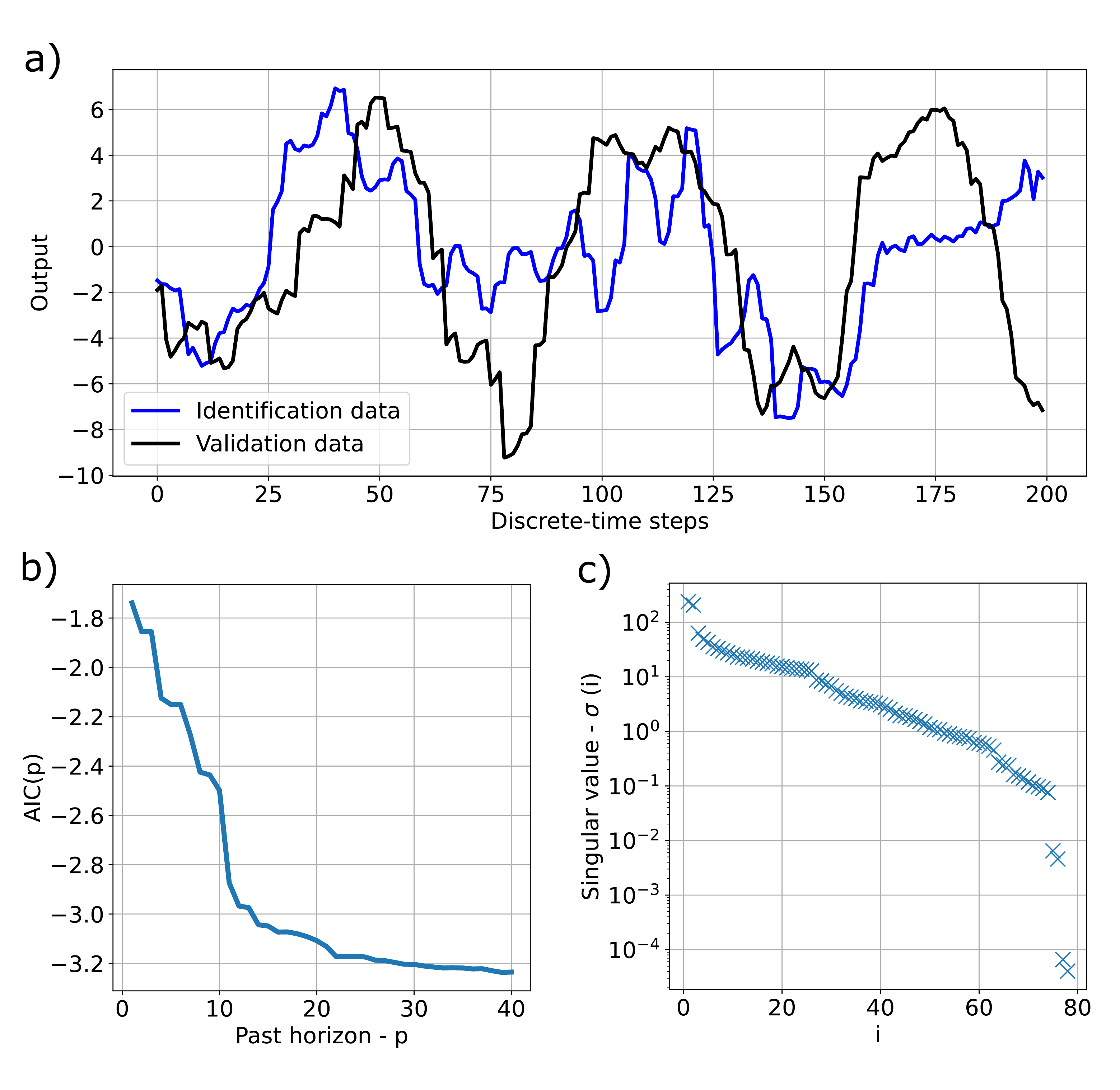}
	\caption{Identification results for disturbances generated by the piezo tip-tilt mirror. (a) Identification and validation time-series used to identify the model. The output is the $x$ projection of the spot center. (b) The AIC value as the function of the past window $p$ used to estimate the Markov parameter matrix. (c) Singular values of the matrix $\mathcal{D}$ defined in \eqref{singularValueDecomposition}.}
	\label{fig:Graph5}
\end{figure}

Figure~\ref{fig:Graph5}(a) shows the identification and validation time series. We use the AIC value to estimate $p$ that defines the Markov matrix $\hat{M}_{p}$ in \eqref{estimateOfMarkovParameters}. The final estimate is the value of $p$ for which the AIC value is smallest, for more details see~\cite{haber2020modelingHaberVerhaegen}. Figure~\ref{fig:Graph5}(b) shows the plot of AIC values. We select the past window of $p=39$. The future window $f$ estimate is $38$. Figure~\ref{fig:Graph5}(c) shows the singular value plot of the matrix $\mathcal{D}$. We can observe that there is a gap in singular values around $i=27$. Consequently, our state order estimate is $n=27$. We can also observe a significant gap around $i=74$. However, this is a large state order that increases the variance of the estimated model. That is, this state estimate overfits the model. Consequently, we selected a smaller state order of $i=27$ to prevent data overfitting.

\begin{figure}[H]
	\centering 
	\includegraphics[scale=0.3,trim=0mm 0mm 0mm 0mm ,clip=true]{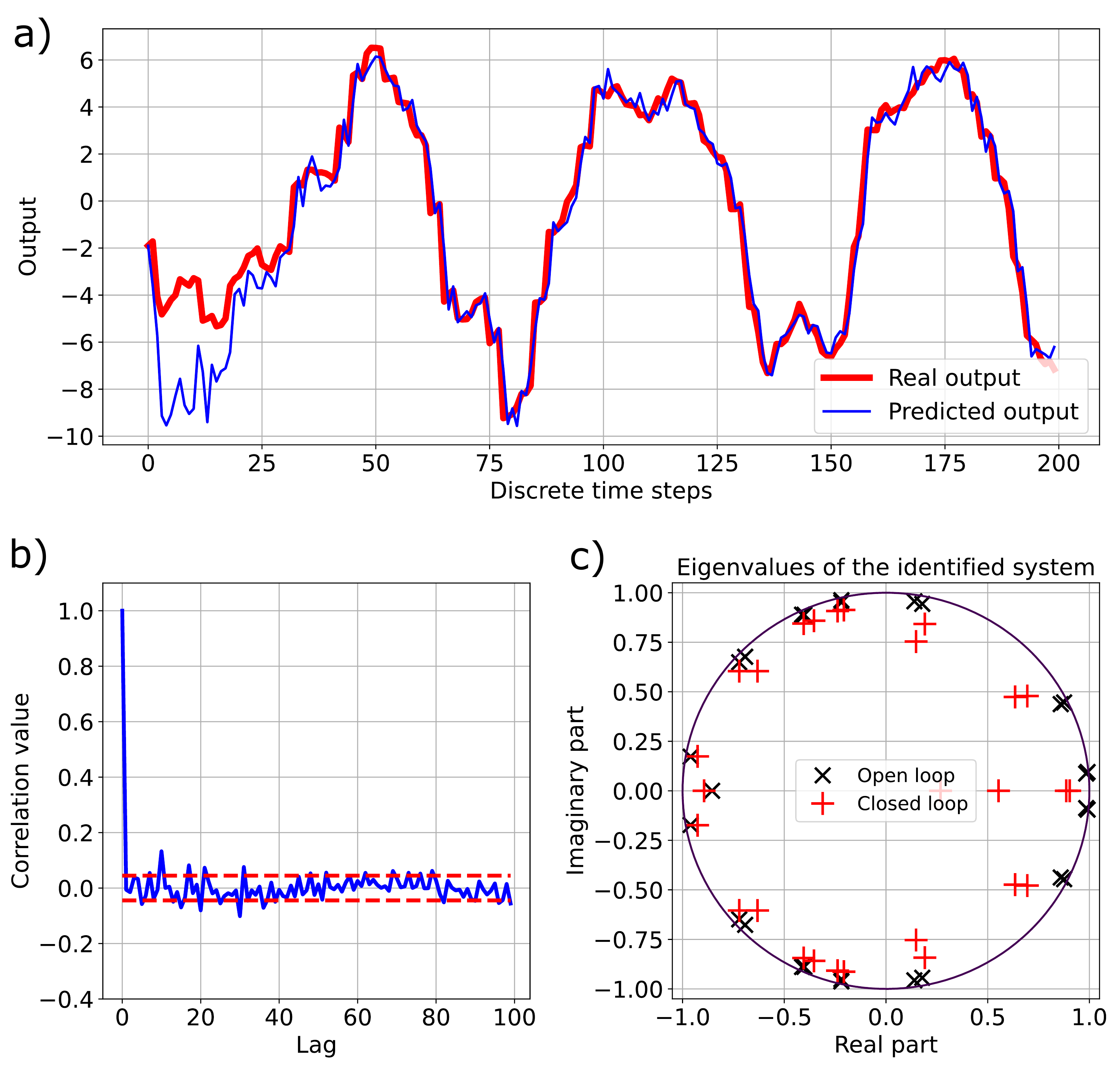}
	\caption{Identification results for disturbances generated by the piezo tip-tilt mirror. (a) The output predicted by the estimated model ("Predicted output") and observed output ("Real output"). (b) The autocorrelation function of the identification error computed on the basis of the predicted output and observed output. (c) The eigenvalues of the estimated matrix $\hat{A}$ ("Open loop") and the eigenvalues of the estimated matrix $\hat{\bar{A}}$ ("Closed loop").}
	\label{fig:Graph6}
\end{figure}

After we estimate the model, we validate the model performance. Figure~\ref{fig:Graph6}(a) shows the output predicted by the model ("Predicted output") and the observed output ("Real output"). We can observe that the identified Kalman filter is able to accurately track the output. Figure~\ref{fig:Graph6}(b) shows the autocorrelation function of the error between the predicted output  and the observed output. In an ideal case, this autocorrelation function should match the autocorrelation function of a white-noise sequence. That is, if the error sequence is a white noise sequence, the autocorrelation value for the lag of $0$ should be equal to $1$, and all other autocorrelation coefficients should be in the region limited by the red dashed lines (see experimental part of Section~\ref{section3} for more details on the white-noise hypothesis test). In our case, $26$ out of $100$ entries are outside the limits. This indicates that our model can still be improved. This can be achieved by changing the combination of the past window, future window, and state order parameters. Despite this Fig.~\ref{fig:Graph6}(a) shows that our model can still accurately track the output. This is also confirmed by the very high value of $97 \%$ of the Variance Accounted For (VAF) of the estimated model. Finally, Fig.~\ref{fig:Graph6}(c) shows the eigenvalues of the estimated matrix $\hat{A}$ (open-loop matrix) and the eigenvalues of the estimated matrix $\hat{\bar{A}}$ (closed-loop matrix). From this eigenvalue plot, we can observe that both open-loop and closed-loop Kalman observer systems are asymptotically stable. This is because all the eigenvalues are inside of the unit circle which is the stability boundary for discrete-time systems. 

\begin{figure}[H]
	\centering 
	\includegraphics[scale=0.3,trim=0mm 0mm 0mm 0mm ,clip=true]{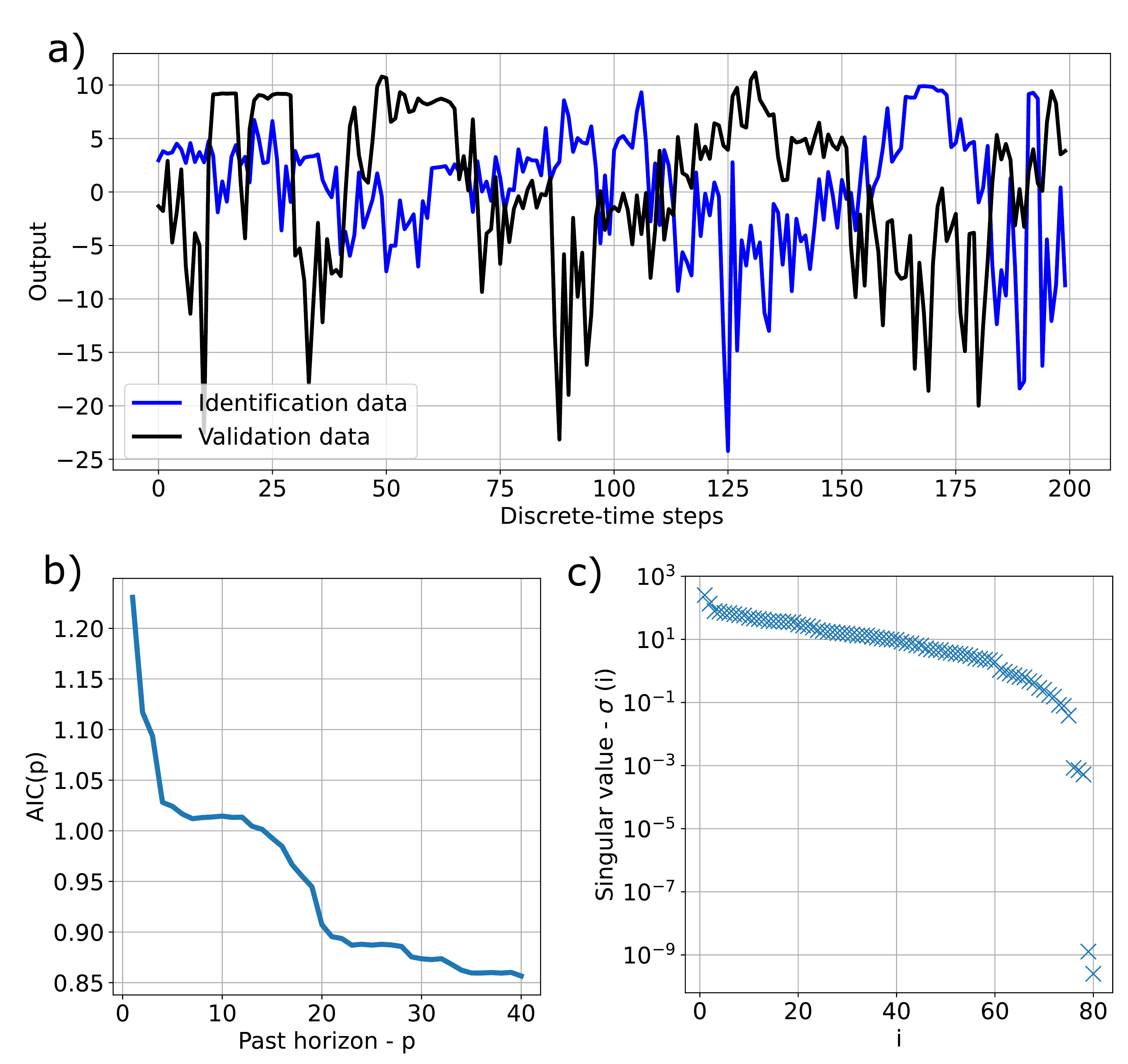}
	\caption{Identification results for disturbances generated by the linear piezo stage.  (a) Identification and validation time-series used to identify the model. The output is the $x$ projection of the spot center. (b) The AIC value as the function of the past window $p$ used to estimate the Markov parameter matrix. (c) Singular values of the matrix $\mathcal{D}$ defined in \eqref{singularValueDecomposition}.}
	\label{fig:Graph7}
\end{figure}

\begin{figure}[H]
	\centering 
	\includegraphics[scale=0.3,trim=0mm 0mm 0mm 0mm ,clip=true]{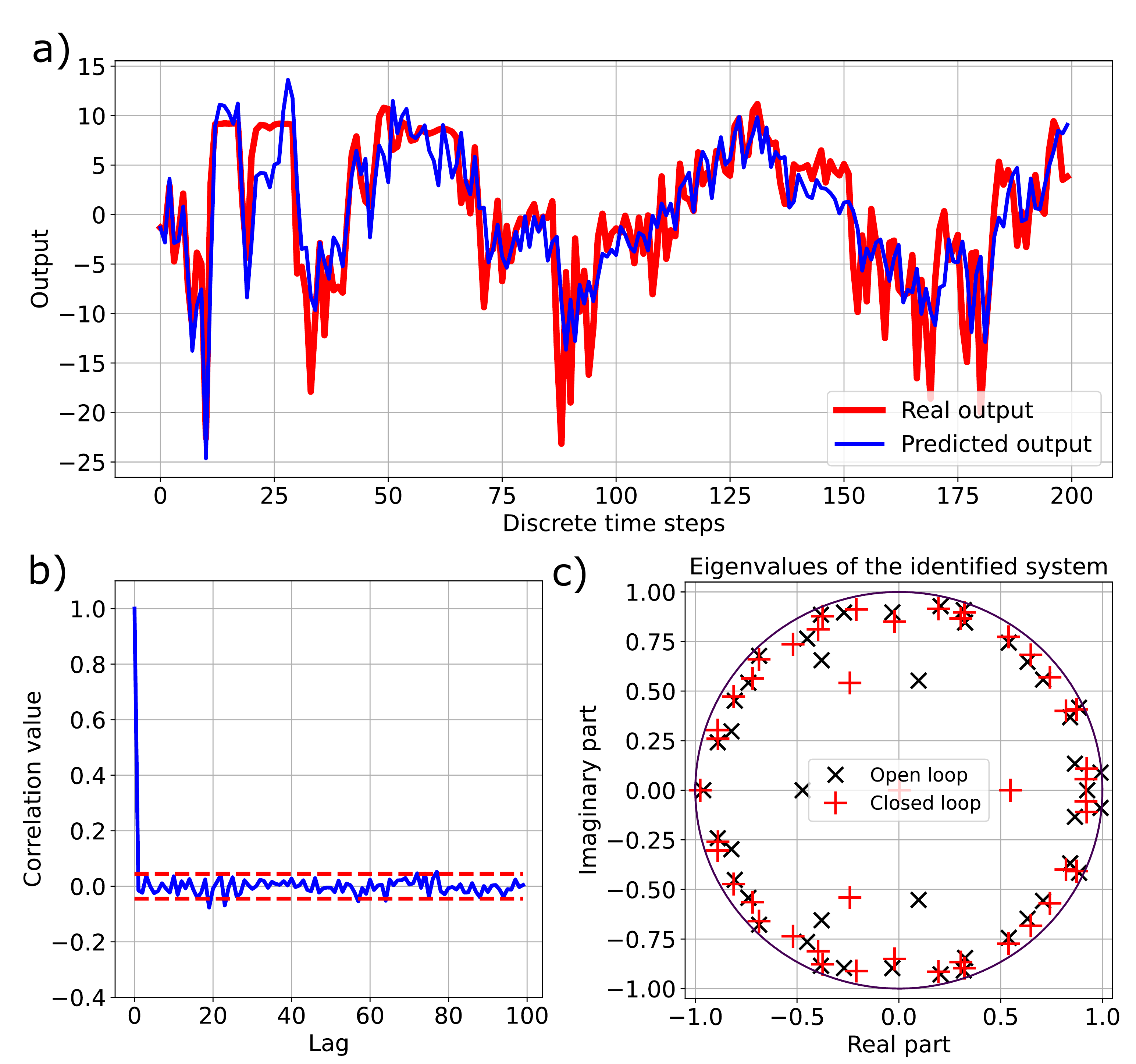}
	\caption{Identification results for disturbances generated by the linear piezo stage.  (a) The output predicted by the estimated model ("Predicted output") and observed output ("Real output"). (b) The autocorrelation function of the identification error computed on the basis of the predicted output and observed output. (c) The eigenvalues of the estimated matrix $\hat{A}$ ("Open loop") and the eigenvalues of the estimated matrix $\hat{\bar{A}}$ ("Closed loop").}
	\label{fig:Graph8}
\end{figure}

Next, we test the subspace identification algorithm on a data set generated by applying the disturbance signals to the linear piezo stage in Fig.~\ref{fig:Graph1}. The system identification results are shown in Fig.~\ref{fig:Graph7} and Fig.~\ref{fig:Graph8}. The figure captions are equivalent to the captions of Figs.~\ref{fig:Graph6} and~\ref{fig:Graph7}. In this case, the estimate parameters are: past window $p=40$, future window $f=23$, and model order of $n=45$. The variance accounted for is $58$. We have $8$ autocorrelation coefficients out of $100$ that exceed the white-noise autocorrelation limits. We can observe that the linear piezo stage induced higher-order dynamics compared to the piezo tip-tilt mirror. This is because the linear piezo stage together with the bracket has less damped dynamics compared to the piezo tip-tilt mirror. The estimated model deliberately does not fit higher-order random oscillations since they decrease the model quality verified on the validation data set.

Overall, in both cases, the identification results are good and clearly demonstrate the great potential of the subspace identification method for directly estimating Kalman filter models of the stochastic spot dynamics.

\section{Conclusions and future work}
\label{section5}

In this manuscript, we developed a unified data-driven Kalman filter approach for covariance estimation and system identification of the stochastic dynamics of the optical spot position. We experimentally demonstrated that after covariance matrices are estimated, approximate first-principle Kalman filter models can be an effective tool for tracking the spot dynamics. Then, we experimentally demonstrated the great potential of the subspace identification methods for directly estimating the Kalman filter models of spot dynamics from the observed time series. In future work, we will investigate and compare the performance of other types of system identification methods for estimating spot dynamics. Also, we will use the framework developed in this paper to develop optimal controllers for suppressing the disturbance dynamics.

\section*{Funding}
This work is funded by the NASA SBIR Phase I grant with the contract number of 80NSSC22PB172.

\section*{Disclosures}

The authors declare no conflict of interests.

\section*{Data availability}
Data underlying the results presented in this paper are not publicly available at this time but may be obtained from the authors upon reasonable request.

\bibliography{sample}
\bibliographystyle{spiebib} 

\end{document}